\documentclass[12pt]{iopart}
\usepackage{graphicx}
\usepackage{iopams}
\usepackage{amsopn}
\usepackage{hyperref}
\usepackage{cite}
\usepackage{bm}
\usepackage{mathdots} 
\def\bra#1{\mathinner{\langle{#1}|}}
\def\ket#1{\mathinner{|{#1}\rangle}}
\def\coloneq{\mathrel{\mathop:}=}
\DeclareMathOperator{\rnk}{rank}
\DeclareMathOperator{\bin}{bin}
\DeclareMathOperator{\diag}{diag}
\newcommand{\Id}{\mathbf{1}}
 \newtheorem{theorem}{Theorem}
 \newtheorem{definition}{Definition}

\begin{document}
\title{Maximally genuine multipartite entangled mixed X-states of $N$-qubits}
\author{Paulo E M F Mendon\c ca$^1$\footnote{Present address:
ARC Centre for Engineered Quantum Systems, School of Mathematics and Physics, The University of Queensland, St. Lucia, Queensland 4072, Australia}, Seyed Mohammad Hashemi Rafsanjani$^2$,
Di\'ogenes Galetti$^3$ and Marcelo A Marchiolli$^4$}

\address{$ˆ1$ Academia da For\c{c}a A\'{e}rea, C.P. 970, 13.643-970 Pirassununga, SP, Brazil}
\address{$ˆ2$ Rochester Theory Center and the Department of Physics \& Astronomy, University of Rochester, Rochester, NY 14627, USA}
\address{$ˆ3$ Instituto de F\'isica Te\'orica, Universidade Estadual Paulista, Rua Dr. Bento Teobaldo Ferraz 271, Bloco II, Barra Funda, 01140-070 S\~{a}o Paulo, SP, Brazil}
\address{$ˆ4$ Avenida General Os\'orio 414, centro, 14.870-100 Jaboticabal, SP, Brazil}
\eads{\mailto{pmendonca@gmail.com}, \mailto{hashemi@pas.rochester.edu}, \mailto{galetti@ift.unesp.br}, \mailto{marcelo\_march@bol.com.br}}
\begin{abstract}
For every possible spectrum of $2^N$-dimensional density operators, we construct an $N$-qubit X-state of same spectrum and maximal genuine multipartite (GM-) concurrence, hence characterizing a global unitary transformation that --- constrained to output X-states --- maximizes the GM-concurrence of an arbitrary input mixed state of $N$-qubits. We also apply semidefinite programming methods to obtain $N$-qubit X-states with maximal GM-concurrence for a given purity and to provide an alternative proof of optimality of a recently proposed set of density matrices for the purpose, the so-called X-MEMS. Furthermore, we introduce a numerical strategy to tailor a quantum operation that converts between any two given density matrices using a relatively small number of Kraus operators. We apply our strategy to design short operator-sum representations for the transformation between any given $N$-qubit mixed state and a corresponding X-MEMS of same purity.
\end{abstract}
\submitto{\jpa}

\section{Introduction}
In the framework of quantum information theory, mixed $N$-qubit X-states synthesize a family of quantum states whose inherent correlations are much easier to quantify than is generally the case. The prefix `X' is motivated by the shape of their density matrices written in the computational basis~\cite{07Yu459}, whose non-zero entries are either diagonal or anti-diagonal (or, otherwise, can be brought to this form via a local unitary (LU) transformation). Owing to this sparse structure, that includes important states (e.g., Bell's~\cite{00Nielsen}, Werner's~\cite{89Werner4277}, isotropic~\cite{99Horodecki4206}, GHZ~\cite{89Greenberger69}, etc.), \emph{analytical} investigations of entanglement properties\cite{06Wang4343,07Yu459,12Quesada1322,10Weinstein032326,12Rafsanjani062303,13Agarwal1350043,14Mendonca79,15Dunkl} and quantum discord~\cite{08Luo042303,10Ali042105,10Ali069902E,10Fanchini052107,11Girolami052108,11Lu012327,11Chen042313,12Vinjanampathy095303,13Huang014302,14Namkung,15Beggi573,14Giorgi,15Trapp} in $N$-qubit X-states have lately become an active and fruitful field of research.

Retrospectively, important members of the class of two-qubit X-states were identified in~\cite{00Ishizaka22310,01Verstraete12316}, where the concept of maximally entangled mixed states (MEMS) was introduced and characterized. From these seminal works, worthy of note is the observation that two-qubit states of a fixed spectrum and maximal entanglement (as measured by concurrence, negativity or relative entropy of entanglement) can always be found in the X-form. Subsequently, Munro \emph{et al.}~\cite{01Munro30302,03Wei22110} characterized two-qubit states of maximal entanglement for a fixed mixedness (as measured by purity, linear entropy or von Neumman entropy), once again obtaining X-states as results.

In spite of these early achievements, to date, little has been accomplished in extending the characterization of MEMS beyond two-qubits. Largely, this is because sensible measures of \emph{genuine multipartite} entanglement have been identified only recently~\cite{11Jungnitsch190502,11Ma062325} and are generally hard to evaluate, let alone maximize.

A first important step toward the identification of $N$-qubit MEMS for $N>2$ was given by Hashemi Rafsanjani \emph{et al}.~\cite{12Rafsanjani062303}, who showed that the GM-concurrence of $N$-qubit \emph{X-states} admits a simple closed formula, amenable to maximization. Although the resulting optimal states of this maximization cannot be guaranteed to be actual MEMS, at least they are provably MEMS amongst all $N$-qubit X-states. Therefore, in~\cite{13Agarwal1350043}, Agarwal and Hashemi Rafsanjani maximized the X-state GM-concurrence formula under the constraint of a fixed linear entropy, determining the so-called X-MEMS.

In this paper, we enlarge the scope of the term X-MEMS to enclose two classes of X-states: \emph{X-MEMS with respect to} (wrt) \emph{spectrum}, referring to those $N$-qubit X-states of maximal GM-concurrence for a fixed spectrum, in analogy to the original MEMS introduced in~\cite{00Ishizaka22310,01Verstraete12316}; and \emph{X-MEMS wrt purity}, referring to $N$-qubit X-states that maximize the GM-concurrence for a fixed value of purity, in parallel with~\cite{01Munro30302,03Wei22110}. Our main results initially consist of: (i) a complete characterization of X-MEMS wrt spectrum, and (ii) a demonstration that X-MEMS wrt purity can be obtained from the solution of a semidefinite program (SDP)~\cite{96Vandenberghe49,04Boyd}, by which means (iii) we provide an alternative proof of optimality of the states obtained in~\cite{13Agarwal1350043}. Moreover, (iv) we characterize the unitary transformation that maximizes the X-state GM-concurrence formula of an arbitrary $N$-qubit state, generalizing the result of~\cite{01Verstraete12316} for $N=2$. Finally, of independent interest (but also relevant in this context), (v) we construct a family of iterated SDPs whose solutions produce quantum operations (CPTP maps) that implement a desired state transformation with a decreasing number of Kraus operators. The method is illustrated with the determination of short operator-sum representations for the conversion between an arbitrary input state of purity $P$ and a corresponding X-MEMS wrt purity $P$.

Our paper is structured as follows. In section~\ref{sec:preliminaries}, we briefly review the concept of GM-concurrence and, in particular, its simple formula for $N$-qubit X-states. In section~\ref{sec:XMEMSwrtSpectrum}, we characterize X-MEMS wrt spectrum and the unitary transformations that produce such states from arbitrary $N$-qubit density matrices. In section~\ref{sec:XMEMSwrtPurity}, X-MEMS wrt purity are constructed and have their optimality established via SDP theory. Section~\ref{sec:convert} outlines the iterated SDP method to design quantum state transformation with few Kraus operators and exemplifies the method while producing X-MEMS wrt purity from arbitrary input states of the same purity. Finally, section~\ref{sec:conclusion} summarizes our results and discusses some possible avenues of future work.

\section{GM-concurrence of $N$-qubit X-states}\label{sec:preliminaries}

In this section, we present the formula for the GM-concurrence of $N$-qubit X-states obtained in~\cite{12Rafsanjani062303}. For the benefit of the reader unfamiliar with the current literature on multipartite entanglement (in particular,~\cite{10Huber210501,11Ma062325,12Rafsanjani062303}), we start by reviewing some key definitions concerning $N$-qubit X-states, GM-entanglement and GM-concurrence.

To begin with, let us introduce some notation. Throughout, ${\sf H}_{d_i}$ denotes the (complex) Hilbert space of dimension $d_i$, whereas $\mathcal{B}({\sf H}_{d_i})$ denotes the set of (bounded) linear operators acting on ${\sf H}_{d_i}$. The set of all possible bipartitions of $\{1,2,\ldots,N\}$ is denoted by $\Gamma$ and a particular bipartition $\{A_{\eta}|B_{\eta}\}$ in $\Gamma$ is denoted by $\Gamma_{\eta}$ (with $\eta$ ranging from $1$ to $2^{N-1}-1$). Partial traces over Hilbert spaces ${\sf H}_{d_i}$ whose labels $i$ belong to $B_{\eta}$ are concisely indicated as $\tr_{B_{\eta}}$.

\begin{definition}
An operator $\bm{\rho}_X\in\mathcal{B}({\sf H}_2\otimes\ldots\otimes{{\sf H}_2})$ represents an \emph{$N$-qubit X-state} if and only if, in the computational basis $\{\ket{\bin i}\}_{i=0}^{2^N-1}$ (and up to a LU-transformation), it assumes the matrix form 
\begin{equation}\label{eq:NqubitXstate}
\fl \bm{\rho}_X=\left[\begin{array}{cccccccc}
a_1    &       &       &       &  &       &   &r_1e^{i\phi_1}\\
       &a_2    &       &       &  &       &r_2e^{i\phi_2} &   \\
       &       &\ddots &       &  &\iddots&   &   \\
       &       &       &a_n    &r_ne^{i\phi_n}&       &   &   \\
       &       &       &r_ne^{-i\phi_n}&b_n&       &   &   \\
       &       &\iddots&       &  &\ddots &   &   \\
       &r_2e^{-i\phi_2}&        &       &  &       &b_2&   \\
r_1e^{-i\phi_1}&       &        &       &  &       &   &b_1
\end{array}\right]\,,
\end{equation}
where $n\coloneq 2^{N-1}$ and, for every integer $k\in[1,n]$, we have $a_k$, $b_k$, $r_k \in \mathbb{R}_+$, $\phi_k\in[0,2\pi]$, 
\begin{equation}\label{eq:normandpsd}
\sum_{k=1}^n (a_k+b_k)=1\quad\mbox{and}\quad 0\leq r_k\leq \sqrt{a_k b_k}\,.
\end{equation}
\end{definition}
While~(\ref{eq:NqubitXstate}) visually justifies the prefix `X', the conditions (\ref{eq:normandpsd}) ensure the normalization and positive semidefiniteness of $\bm{\rho}_X$. As a glance at~(\ref{eq:NqubitXstate}) demonstrates, the index $k\in[1,n]$ can be regarded as a label for uncoupled bidimensional subspaces. That any $N$-qubit X-state is decomposable into $n$ such subspaces is a key property that will be implicitly exploited throughout this paper. 

Although we are only interested in the entanglement properties of $N$-qubit X-states, we proceed with a general definition of GM-entanglement.

\begin{definition}\label{def:GMent}
An $N$-partite density operator $\bm{\rho}\in\mathcal{B}({\sf H}_{d_1}\otimes {\sf H}_{d_2}\otimes\ldots\otimes {\sf H}_{d_N})$ is \emph{GM-entangled} if and only if it is not \emph{biseparable}.
\end{definition}
To understand the concept of biseparability, consider first its definition for pure states.

\begin{definition}\label{def:biseppure}
An $N$-partite state $\ket{\psi}\in{\sf H}_{d_1}\otimes {\sf H}_{d_2}\otimes\ldots \otimes{\sf H}_{d_N}$ is \emph{biseparable} if and only if there is a Hilbert space bipartition ${\sf H}_A\otimes {\sf H}_B = {\sf H}_{d_1}\otimes {\sf H}_{d_2}\otimes\ldots \otimes{\sf H}_{d_N}$ and a pair of states $\ket{\psi_A}\in{\sf H}_A$, $\ket{\psi_B}\in{\sf H}_B$, such that $\ket{\psi}=\ket{\psi_A}\otimes\ket{\psi_B}$.
\end{definition}
Note that definition~\ref{def:biseppure} implies that a biseparable state is not necessarily separable, as there might be entanglement within ${\sf H}_A$ and/or ${\sf H}_B$. It then follows from definition~\ref{def:GMent}, that the condition for GM-entanglement is generally more stringent than the condition for bipartite entanglement, for example. In fact, GM-entanglement only occurs when bipartite entanglement is observed across all possible bipartitions of ${\sf H}_{d_1}\otimes {\sf H}_{d_2}\otimes\ldots \otimes{\sf H}_{d_N}$.

The notion of biseparability is extended to mixed states as follows.

\begin{definition}
An $N$-partite density operator $\bm{\rho}\in\mathcal{B}({\sf H}_{d_1}\otimes {\sf H}_{d_2}\otimes\ldots \otimes{\sf H}_{d_N})$ is \emph{biseparable} if and only if it can be decomposed in an ensemble of biseparable pure states, that is
\begin{equation}
\bm{\rho}=\sum_i p_i \ket{\psi_i}\!\bra{\psi_i}\,,
\end{equation}
where $\sum_i p_i=1$ and each $\ket{\psi_i}$ is biseparable (even if with respect to different bipartitions of ${\sf H}_{d_1}\otimes {\sf H}_{d_2}\otimes\ldots \otimes{\sf H}_{d_N}$).
\end{definition}

The above definitions provide a formal criterion to determine whether a general mixed state is GM-entangled or not. A further step was given by Ma \emph{et al.}~\cite{11Ma062325}, who introduced the GM-entanglement measure named GM-concurrence.

\begin{definition} The GM-concurrence of an $N$-partite pure state $\ket{\psi}\in{\sf H}_{d_1}\otimes {\sf H}_{d_2}\otimes\ldots \otimes{\sf H}_{d_N}$ is given by
\begin{equation}
C_{\rm GM}(\ket{\psi})\coloneq \min_{\eta\in\{1,\ldots, 2^{N-1}-1\}}\sqrt{2}\sqrt{1-\tr[\bm{\rho}^2_{A_{\eta}}]}\,,
\end{equation}
where $\bm{\rho}_{A_{\eta}}\coloneq \tr_{B_{\eta}}[\ket{\psi}\!\bra{\psi}]$. For $N$-partite density operators $\bm{\rho}\in\mathcal{B}({\sf H}_{d_1}\otimes {\sf H}_{d_2}\otimes\ldots \otimes{\sf H}_{d_N})$, the GM-concurrence is obtained via the convex roof construction
\begin{equation}
C_{\rm GM}(\bm{\rho})= \inf_{\{p_i,\ket{\psi_i}\}}\sum_i p_i C_{\rm GM}(\ket{\psi_i})\,,
\end{equation}
with the infimum taken over all possible ensembles $\{p_i,\ket{\psi_i}\}$ that realize $\bm{\rho}$.
\end{definition}
The GM-concurrence takes its name from the fact that, in the case of two-qubit systems, it matches the Wootters concurrence~\cite{98Wootters2245} and, more generally,  can be shown~\cite{11Ma062325} to satisfy the following minimal requirements for any GM-entanglement measure:
\begin{itemize}
\item GM-entanglement detection: 
\begin{equation}
C_{\rm GM}(\bm{\rho})\geq 0\,,
\end{equation}
 with saturation if and only if $\bm{\rho}$ is biseparable.
\item Convexity: 
\begin{equation}
C_{\rm GM}\left(\sum_i p_i \bm{\rho}_i\right)\leq \sum_i p_i C_{\rm GM}(\bm{\rho}_i)\,.
\end{equation}
\item Monotonicity under local operations and classical communication ($\Omega_{\rm LOCC}$):
\begin{equation}
C_{\rm GM}(\Omega_{\rm LOCC}[\bm{\rho}])\leq C_{\rm GM}(\bm{\rho})\,.
\end{equation}

\item Invariance under LU-transformations ($U_{\rm L}$):
\begin{equation}
C_{\rm GM}(U_{\rm L}\bm{\rho}U_{\rm L}^\dagger)= C_{\rm GM}(\bm{\rho})\,.
\end{equation}
\end{itemize}

Though well motivated, $C_{\rm GM}(\bm{\rho})$ is generally hard to evaluate due to the infimum over all ensembles that realize $\bm{\rho}$. To alleviate this problem, the authors of~\cite{11Ma062325} relied on certain sufficient criteria for GM-entanglement detection proposed by Huber \emph{et al.}~\cite{10Huber210501} to determine computable lower bounds for $C_{\rm GM}$. In particular, if the main- and anti-diagonal entries of $\bm{\rho}$ are parametrized as in~(\ref{eq:NqubitXstate}) (the remaining entries being arbitrary), then one of Ma's lower bounds reads (see~\cite[Appendix A]{15Mendonca} for an explicit derivation)
\begin{equation}
C_{\rm GM}(\bm{\rho})\geq 2\max\left\{0,\max_{k\in [1,n]}\left[r_k-\sum_{j\neq k}^n\sqrt{a_jb_j}\right]\right\}\,.
\end{equation}
Remarkably, as shown by Hashemi Rafsanjani \emph{et al.}~\cite{12Rafsanjani062303}, this lower bound is saturated when $\bm{\rho}=\bm{\rho}_X$, namely,
\begin{equation}\label{eq:CgeNqubitXstate}
C_{\rm GM}(\bm{\rho}_X)= 2\max\left\{0,\max_{k\in [1,n]}\left[r_k-\sum_{j\neq k}^n\sqrt{a_jb_j}\right]\right\}\,.
\end{equation}

The fact that $N$-qubit X-states have their GM-concurrence expressed as a closed formula cannot be overstated. It contrasts with the great difficulty involved in merely detecting GM-entanglement in more general systems, not to mention quantifying it. Of course, this result becomes even more appealing when one notices that $N$-qubit states of practical interest do occur in the X-form (see, e.g., \cite{14Giampaolo93033}), or otherwise can usually be well approximated to it via LU-transformations~\cite{15Mendonca}. Finally, it is interesting that for GHZ-diagonal states (X-states with $a_i=b_i$), the value of GM-concurrence is proportional to the distance of the GHZ-state to the set of biseparable states~\cite{13Rafsanjani062331}.

\section{X-MEMS with respect to spectrum}\label{sec:XMEMSwrtSpectrum}

As mentioned before, the two-qubit MEMS with a given spectrum, characterized in~\cite{00Ishizaka22310,01Verstraete12316}, are X-states. In this section, we \emph{assume} that this is also true in the $N$-qubit case ($N>2$), and characterize the ``$N$-qubit MEMS'' resulting from this assumption. Since it is not known in which circumstances the restriction to the set of X-states is an active constraint\footnote{By an \emph{active constraint} we mean a restriction that is not satisfied unless it is explicitly imposed.} for $N>2$, we adopt the nomenclature introduced in~\cite{13Agarwal1350043} and talk about X-MEMS instead of simply MEMS. 

The results of this section are summarized in Theorem~\ref{thm:XMEMSwrtSpectrum}, which is deliberately presented in close resemblance to the statement of the related Theorem presented in~\cite{01Verstraete12316}, regarding the case $N=2$. The proofs, however, are established in very different ways.

\begin{theorem}\label{thm:XMEMSwrtSpectrum}
The maximal GM-concurrence attainable by an $N$-qubit X-state of spectrum $\bm{\Lambda}$, determined by the eigenvalues $\lambda_1\geq \lambda_2 \geq \ldots \geq \lambda_{2n}$ (with $n\coloneq 2^{N-1}$), is given by
\begin{equation}\label{eq:Cgmoptimal}
\max\left[0,\lambda_{1}-\lambda_{n+1}-2\sum_{\ell = 2}^n\sqrt{\lambda_\ell\lambda_{2n+2-\ell}}\right]\,.
\end{equation}
Moreover, any $N$-qubit density matrix $\bm{\rho_\Lambda}$ (of spectrum $\bm{\Lambda}$) can be coherently transformed into $\bm{\rho_\Lambda}^\prime$, an isospectral $N$-qubit X-density matrix of maximal GM-concurrence, according to $\bm{\rho_\Lambda}^\prime=\bm{\mathcal{U}}\bm{\rho_\Lambda}\bm{\mathcal{U}}^\dagger$, with the unitary $\bm{\mathcal{U}}$ given by
\begin{equation}\label{eq:globalunitary}
\bm{\mathcal{U}}=\left(\bigotimes_{k=1}^N \bm{U}_k\right) \left[\begin{array}{c|c}
\bm{V}_{11} & \bm{V}_{12}\\\hline
\bm{V}_{21} & \bm{V}_{22}
\end{array}\right] \bm{D}_\phi \bm{\Phi}^\dagger\,.
\end{equation}
In~(\ref{eq:globalunitary}), the following definitions apply: $\{\bm{U}_k\}_{k=1}^N$ is a set of arbitrary single qubit unitary operations, $\bm{D}_\phi$ is an arbitrary unitary and diagonal matrix, $\bm{\Phi}$ is the unitary matrix formed from the eigenvectors of $\bm{\rho_\Lambda}$ (such that $\bm{\rho_\Lambda}=\bm{\Phi}\bm{\Lambda}\bm{\Phi}^\dagger$), and 
\begin{equation}
\fl\bm{V}_{11}=\bm{V}_{12}+\sum_{i=2}^n \bm{E}_{i,i}\,,\;\,\bm{V}_{12}=\frac{1}{\sqrt{2}}\bm{E}_{1,1}\,,\;\, \bm{V}_{21}=\frac{1}{\sqrt{2}}\bm{E}_{n,1}\,,\;\,
\bm{V}_{22}=-\bm{V}_{21}+\sum_{i=1}^{n-1} \bm{E}_{i,i+1}\,.\label{eq:V}
\end{equation}
Here, $\bm{E}_{i,j}$ is the $n$-dimensional matrix whose only non-zero entry is equal to $1$ and occupies the $i$th row and $j$th column.
\end{theorem}

An immediate remark is that, as expected, both the optimal GM-concurrence~(\ref{eq:Cgmoptimal}) and the unitary transformation (\ref{eq:globalunitary}) reduce to the corresponding expressions in~\cite[Theorem 1]{01Verstraete12316} when $N=2$. The remainder of this section is devoted to proving the theorem for $N>2$. Essentially, our proof consists of a direct maximization of the GM-concurrence formula of $N$-qubit X-states under the constraint of a fixed spectrum.

Let $\bm{\rho}_{\tilde{\bm{\Lambda}}}^\prime$ denote a generic $N$-qubit X-density matrix of spectrum $\tilde{\bm{\Lambda}}$. We start by taking matrix (\ref{eq:NqubitXstate}) as a parametrization for $\bm{\rho}_{\tilde{\bm{\Lambda}}}^\prime$  and writing a general formula for $C_{\rm GM}(\bm{\rho}_{\tilde{\bm{\Lambda}}}^\prime)$ in terms of its eigenvalues
\begin{equation}\label{eq:eigslambdak}
\lambda_k^\pm=\frac{a_k+b_k}{2}\pm\sqrt{r_k^2+d_k^2}\quad\mbox{for every}\quad k\in[1,n]\,.
\end{equation}
Here, $\lambda_k^\pm$ denote the greatest ($+$) and smallest ($-$) eigenvalues of $\bm{\rho}_{\tilde{\bm{\Lambda}}}^\prime$ associated with the bidimensional subspace labelled by $k$, and $d_k\coloneq (b_k-a_k)/2$. It follows trivially from (\ref{eq:eigslambdak}) that,
\begin{equation}\label{eq:mixabreigs}
\sqrt{r_{k}^2+d_{k}^2}=\frac{\lambda_{k}^+-\lambda_{k}^-}{2}\quad\mbox{and}\quad \sqrt{a_j b_j- r_j^2}= \sqrt{\lambda_j^+ \lambda_j^-}\,,
\end{equation}
which used in (\ref{eq:CgeNqubitXstate}) yields
\begin{equation}\label{eq:eigsCgeNqubitXstate}
C_{\rm GM}(\bm{\rho}_{\tilde{\bm{\Lambda}}}^\prime)=2\max\left[0,\sqrt{\left(\frac{\lambda_{1}^+-\lambda_{1}^-}{2}\right)^2-d_1^2}-\sum_{j=2}^n\sqrt{\lambda_j^+\lambda_j^- +r_j^2}\right],
\end{equation}
where, without loss of generality, we fixed the label $k=1$ to the subspace whose value of $r_k-\sum_{j\neq k}^n\sqrt{a_j b_j}$ is the largest.

Our goal now is to maximize (\ref{eq:eigsCgeNqubitXstate}) under the contraint that the set of eigenvalues $\{\lambda_j^\pm\}_{j=1}^n$ matches the set of eigenvalues of the arbitrary (but given) $N$-qubit density matrix $\bm{\rho_\Lambda}$, i.e., $\tilde{\bm{\Lambda}}=\bm{\Lambda}$. To do that, let us first consider the optimization over $d_1$ and $\{r_j\}_{j=2}^n$. Although these variables are constrained by (\ref{eq:normandpsd}) and related to $\{\lambda_j^\pm\}_{j=1}^n$ by (\ref{eq:eigslambdak}), we will momentarily ignore these contraints. By doing so, we significantly simplify the optimization procedure at the expense of risking over-maximization $C_{\rm GM}$. Nevertheless, as we will soon demonstrate, the resulting maximal is actually attainable, meaning that our simplifying assumptions are harmless. With that in mind we set, for every $j\in[2,n]$,
\begin{equation}\label{eq:rjd1}
r_j=d_1=0\,,
\end{equation}
which clearly maximizes (\ref{eq:eigsCgeNqubitXstate}) over $r_j$ and $d_1$. 

At this point, we are left with the maximization over $\{\lambda_j^\pm\}_{j=1}^n$, written as
\begin{equation}\label{eq:optprblm1}
\mbox{maximize}\quad\lambda_{1}^+-\bm{u}\cdot\bm{v}\quad\mbox{subject to}\quad \{\lambda_j^\pm\}_{j=1}^n=\{\lambda_j\}_{j=1}^{2n}\,,
\end{equation}
where $\lambda_1\geq\ldots\geq\lambda_{2n}$ are the eigenvalues of the arbitrary (but given) $N$-qubit density matrix $\bm{\rho_\Lambda}$ and the vectors $\bm{u},\bm{v}\in\mathbb{R}^{2n-1}$ are given by
\begin{eqnarray}
\bm{u}&\coloneq\left(\sqrt{\lambda_{2}^-}\,,\,\ldots\,,\,\sqrt{\lambda_{n}^-}\,,\,\sqrt{\lambda_{1}^-}\,,\,\sqrt{\lambda_{n}^+}\,,\,\ldots\,,\,\sqrt{\lambda_{2}^+}\right)\,,\\
\bm{v}&\coloneq\left(\sqrt{\lambda_{2}^+}\,,\,\ldots\,,\,\sqrt{\lambda_{n}^+}\,,\,\sqrt{\lambda_{1}^-}\,,\,\sqrt{\lambda_{n}^-}\,,\,\ldots\,,\,\sqrt{\lambda_{2}^-}\right)\,.
\end{eqnarray}
Here, we aim to assign to each variable in $\{\lambda_j^\pm\}_{j=1}^n$ an eigenvalue of $\bm{\rho_\Lambda}$, in such a way that $\lambda_1^+$ is maximal and $\bm{u}\cdot\bm{v}$ is minimal. To maximize $\lambda_1^+$, we simply assign to it the largest eigenvalue of $\bm{\rho_\Lambda}$, i.e., 
\begin{equation}\label{eq:lambda1peqlambda1}
\lambda_1^+=\lambda_1\,.
\end{equation}
To minimize $\bm{u}\cdot\bm{v}$, first notice that $\bm{u}$ and $\bm{v}$ display the same entries in the reversed order, with $\sqrt{\lambda_1^-}$ occupying the central position in both vectors. It follows from the rearrangement inequality~(see, e.g., \cite[Theorem 368, page 261]{34Hardy}) that the scalar product between two vectors defined up to the ordering of their entries is minimized if and only if they are sorted in opposite directions. So, we make the entries of $\bm{u}$ and $\bm{v}$ monotonically increasing and decreasing, respectively, by assigning, for every $j\in[2,n]$,
\begin{equation}\label{eq:ansatzdistrib}
\lambda_j^+=\lambda_j\,,\quad\lambda_1^-=\lambda_{n+1}\,,\quad\mbox{and}\quad\lambda_j^-=\lambda_{2n+2-j}\,.
\end{equation}

Substituting the identities~(\ref{eq:rjd1}), (\ref{eq:lambda1peqlambda1}) and (\ref{eq:ansatzdistrib}) in (\ref{eq:mixabreigs}) and solving the resulting system for $r_k$, $a_k$ and $b_k$ (under the constraints described in (\ref{eq:normandpsd})), we obtain that\footnote{As a matter of fact, many other solutions can be obtained by interchanging the values of $a_j$ and $b_j$ indicated in (\ref{eq:solsystem}) for any $j\in[2,n]$. However, this does not lead to essentially new X-MEMS, since the X-MEMS corresponding to these solutions can always be generated from the X-MEMS corresponding to (\ref{eq:solsystem}) via a LU-transformation.}, for every $j\in[2,n]$,
\begin{equation}\label{eq:solsystem}
\fl r_1=\frac{\lambda_1-\lambda_{n+1}}{2}\,,\quad a_1=b_1=\frac{\lambda_1-\lambda_{n+1}}{2}\,,\quad r_j=0\,,\quad a_j=\lambda_j\quad\mbox{and}\quad b_j=\lambda_{2n+2-j}\,.
\end{equation}
By plugging (\ref{eq:solsystem}) into (\ref{eq:CgeNqubitXstate}), it is easily seen that (\ref{eq:Cgmoptimal}) holds. In order to see that (\ref{eq:Cgmoptimal}) is physically attainable, substitute (\ref{eq:solsystem}) into matrix (\ref{eq:NqubitXstate}) (and set $\phi_k=0$ for every $k\in[1,n]$), to get
\begin{equation}\label{eq:rhoxprime}
\fl \bm{\rho_\Lambda}^\prime=\frac{1}{2}\left[\begin{array}{cccccccc}
\lambda_1+\lambda_{n+1}    &       &       &       &  &       &   &\lambda_1-\lambda_{n+1}\\
       &2\lambda_2    &       &       &  &       &0 &   \\
       &       &\ddots &       &  &\iddots&   &   \\
       &       &       &2\lambda_n    &0&       &   &   \\
       &       &       &0&2\lambda_{n+2}&       &   &   \\
       &       &\iddots&       &  &\ddots &   &   \\
       &0&        &       &  &       &2\lambda_{2n}&   \\
\lambda_1-\lambda_{n+1}&       &        &       &  &       &   &\lambda_1+\lambda_{n+1}
\end{array}\right]\,.
\end{equation}
It is immediate to check that $\bm{\rho_\Lambda}^\prime$ is a valid X-density matrix with the same spectrum of $\bm{\rho_\Lambda}$ and GM-concurrence given by~(\ref{eq:Cgmoptimal}).

Finally, let us establish (\ref{eq:globalunitary}). Since $\bm{\rho_\Lambda}$ and $\bm{\rho_\Lambda}^\prime$ are isospectral, we can write
\begin{equation}\label{eq:rhoandrhoprime}
\bm{\rho_\Lambda}=\bm{\Phi}\bm{\Lambda}\bm{\Phi}^\dagger\quad\mbox{and}\quad\bm{\rho_\Lambda}^\prime=\bm{V}\bm{\Lambda}\bm{V}^\dagger\,,
\end{equation}
where $\bm{\Phi}$ is the matrix of the eigenvectors of $\bm{\rho_\Lambda}$ and $\bm{V}$ is the matrix of the eigenvectors of $\bm{\rho_\Lambda}^\prime$ given by (\ref{eq:rhoxprime}). Some simple linear algebra shows that, for $\bm{\Lambda}=\diag[\lambda_1,\lambda_2,\ldots,\lambda_{2n}]$, the matrix $\bm{V}$ admits the block decomposition specified in (\ref{eq:V}). Thus, combining the two identities in (\ref{eq:rhoandrhoprime}) to eliminate $\bm{\Lambda}$, we arrive at
\begin{equation}
\bm{\rho_\Lambda}^\prime= \bm{\mathcal{U}} \bm{\rho_\Lambda} \bm{\mathcal{U}}^\dagger\,,\quad \mbox{where}\quad \bm{\mathcal{U}}=\left[\begin{array}{c|c}
\bm{V}_{11} & \bm{V}_{12}\\\hline
\bm{V}_{21} & \bm{V}_{22}
\end{array}\right] \bm{\Phi}^\dagger\,.
\end{equation}
As noted before, the X-MEMS of (\ref{eq:rhoxprime}) are unique up to LU-transformations, for which reason, in (\ref{eq:globalunitary}), the expression of $\bm{\mathcal{U}}$ appears pre-multiplied by an arbitrary LU-transformation. Furthermore, for sake of generality, we have also multiplied an arbitrary (generally non-local) diagonal unitary matrix $\bm{D}_\phi$ in (\ref{eq:globalunitary}). It should be emphasized, though, that $\bm{D}_\phi$ has obviously no effect on the output state $\bm{\rho_\Lambda}^\prime$.

Let us conclude this section by answering a central question that arises from the present work: are $N$-qubit X-MEMS actual $N$-qubit MEMS? Although this is long known to be the case for $N=2$~\cite{01Verstraete12316}, indications that the same may also hold for $N=3$ have only recently appeared in the work of Hedemann~\cite{13Hedemann}. Alas, to the best of our knowledge, the topic seems to be utterly unexplored for $N\geq 4$. To see that that conjecture cannot hold in general, note that an affirmative answer (combined with (\ref{eq:Cgmoptimal})), would imply that $N$-qubit density matrices whose eigenvalues $\lambda_1\geq \lambda_2 \geq \ldots \geq \lambda_{2n}$ satisfy
\begin{equation}\label{eq:sfs}
\lambda_{1}\leq\lambda_{n+1}+ 2\sum_{\ell = 2}^n\sqrt{\lambda_\ell\lambda_{2n+2-\ell}}
\end{equation}
cannot acquire GM-entanglement by means of a global unitary transformation. However, as recently shown by Huber \emph{et al.}~\cite{14Huber}, $N$-qubit thermal states of arbitrarily high temperatures and $N$ sufficiently large (represented, in the computational basis, by diagonal density matrices arbitrarily close to the identity, hence fulfilling (\ref{eq:sfs})) \emph{can} acquire GM-entanglement by means of rotations to Dicke-like (non-X) states, thus providing a counter-example to the original conjecture. 

\section{X-MEMS with respect to purity}\label{sec:XMEMSwrtPurity}

Since the purity of a given quantum state can be expressed as the sum of the squares of their eigenvalues, there are, in general, many spectra that realize a fixed value of purity $P$. In this section, we consider the problem of searching for a maximizer of the $N$-qubit X-state GM-concurrence formula amongst all $N$-qubit spectra that realize $P$. Any X-state with such an optimal spectrum is referred to as an X-MEMS wrt purity, denoted by $\bm{\rho}_P^\prime$.

As a matter of fact, $N$-qubit X-MEMS wrt purity have been recently determined in the work of Agarwal and Hashemi Rafsanjani~\cite{13Agarwal1350043}, whose main findings are summarized in the statement of the following theorem.
\begin{theorem}[Agarwal and Hashemi Rafsanjani]\label{thm:XMEMSwrtPurity}
For any value of purity $P\in ]1/(n+1),1]$ with $n\coloneq 2^{N-1}$, the maximal GM-concurrence attainable by an $N$-qubit X-state of purity $P$ is $2\gamma$, where the parameter $\gamma\in\,]0,1/2]$ is determined by $P$ according to 
\begin{equation}\label{eq:gammafxnP}
\gamma\coloneq\left\{\begin{array}{ll}
\sqrt{\frac{P}{2}-\frac{1}{2(n+1)}}&\mbox{if}\quad \frac{1}{n+1}< P\leq \frac{n+3}{(n+1)^2}\\[3mm]
\frac{1}{2n}+\frac{1}{2}\sqrt{(1-\frac{1}{n})(P-\frac{1}{n})}&\mbox{if}\quad \frac{n+3}{(n+1)^2}\leq P\leq 1\\
\end{array}
\right.\,.
\end{equation}
Up to LU-transformations, every $N$-qubit X-density matrix of purity $P$ that achieves maximal GM-concurrence is given by (\ref{eq:rhoxprime}) with
\begin{equation}\label{eq:propprimalsol}
\lambda_1=f(\gamma)+\gamma\,,\quad \lambda_{j}= g(\gamma)\,,\quad \lambda_{n+1}=f(\gamma)-\gamma\quad\mbox{and}\quad \lambda_{n+j}=0
\end{equation}
for every $j\in[2,n]$, being $f$ and $g$ defined as follows:
\begin{equation}\label{eq:defsfandg}
f(\gamma)\coloneq\left\{\begin{array}{cl}
\frac{1}{n+1}&\mbox{if}\quad 0< \gamma\leq \frac{1}{n+1}\\[2mm]
\gamma&\mbox{if}\quad \frac{1}{n+1}\leq \gamma\leq \frac{1}{2}\\
\end{array}
\right.\quad\mbox{and}\quad
g(\gamma)\coloneq\frac{1-2f(\gamma)}{n-1}\,.
\end{equation}
\end{theorem}

In what follows, we give an alternative proof of this theorem by exploiting the results of Theorem~\ref{thm:XMEMSwrtSpectrum} and the theory of semidefinite programming~\cite{96Vandenberghe49,04Boyd}.

Since purity is determined by the spectrum (and not the other way around), the specification of a spectrum generally represents a stronger constraint than a specification of purity. As a result, if $\bm{\rho}_P^\prime$ is a \emph{X-MEMS wrt purity $P$} and has spectrum $\bm{\Lambda}_P$, then $\bm{\rho}_P^\prime$ is also a \emph{X-MEMS wrt to the spectrum $\bm{\Lambda}_P$}. In other words, every X-MEMS wrt purity can be regarded as a X-MEMS wrt \emph{some} spectrum, in which case every $\bm{\rho}_P^\prime$ is of the form (\ref{eq:rhoxprime}) up to a LU-transformation. Thanks to this, we can determine $\bm{\rho}_P^\prime$ by maximizing the GM-concurrence of (\ref{eq:rhoxprime}) over all sets of physical eigenvalues that realize $P$, which yields the optimization problem\footnote{Note that here $\{\lambda_k\}_{k=1}^n$ represents the set of optimization variables, as opposed to a fixed set of eigenvalues (as it was the case in Sec.~\ref{sec:XMEMSwrtSpectrum}). }
\begin{eqnarray}\label{eq:optprblm2}
\fl\mbox{maximize}&\quad\lambda_{1}-\lambda_{n+1}-\sum_{j=2}^n\sqrt{\lambda_j\lambda_{2n+2-j}}\nonumber\\
\fl\mbox{subject to}&\quad \lambda_1\geq \lambda_2\geq \ldots \geq \lambda_{2n}\geq 0\,,\quad\sum_{k=1}^{2n}\lambda_k=1\quad\mbox{and}\quad \sum_{k=1}^{2n}\lambda_k^2=P\,.
\end{eqnarray}

Next, we give a few arguments that allow some simplification of problem~(\ref{eq:optprblm2}). First, there is no need to explicitly require the ordering $\lambda_1\geq \lambda_2\geq \ldots \geq \lambda_{2n}$ since we know in advance (rearrangement inequality) that this particular ordering\footnote{Up to (irrelevant) permutations of the form $\lambda_j\leftrightarrow\lambda_{2n+2-j}$ for any $j\in[2,n]$.}  is a necessary condition for the maximization of the considered objective function and will be thus satisfied anyway. Second, the objective function is clearly maximized if we set $\lambda_{2n+2-j}=0$ for every $j\in[2,n]$, which does not violate any problem constraint and reduces the equality constraints to 
\begin{equation}
\lambda_{n+1}=1-\sum_{k=1}^{n}\lambda_k\quad\mbox{and}\quad \left(1-\sum_{k=1}^{n}\lambda_k\right)^2+\sum_{k=1}^{n}\lambda_k^2= P\,.
\end{equation}
Given these two points, we can replace the original (non-linear) objective function with the linear function $\lambda_1-\lambda_{n+1}$, and all the inequality constraints with a single one: $\lambda_{n+1}\geq 0$. Finally, without altering the solution of the problem, we can replace the equality in the quadratic constraint with the inequality `$\leq$', thus enlarging the set of feasible points to its convex hull~\cite[Chapter 32]{70TyrrelRockafellar}.  Accordingly, we end up with the equivalent optimization problem on $n$ real variables:
\begin{eqnarray}\label{eq:optprblm3}
\fl\mbox{maximize}&\quad-1+2 \lambda_{1}+\sum_{k=2}^{n}\lambda_k\nonumber\\
\fl\mbox{subject to}&\quad \sum_{k=1}^{n}\lambda_k\leq 1\quad\mbox{and}\quad \left(1-\sum_{k=1}^{n}\lambda_k\right)^2+\sum_{k=1}^{n}\lambda_k^2\leq P\,.
\end{eqnarray}

As explained in~\ref{app:quadtoLMI}, the quadratic (convex) constraint in (\ref{eq:optprblm3}) can be written as a linear matrix inequality (LMI), turning problem~(\ref{eq:optprblm3}) into the SDP~(\ref{eq:SDPopennotation}). It is straightforward to see that (\ref{eq:SDPopennotation}) admits the \emph{SDP standard inequality form}\footnote{It should be noted that, although problems (\ref{eq:SDPopennotation}) and (\ref{eq:optprblm4}) are solved by the same set of eigenvalues, the resulting optimal values of the two problems are not exactly the same. That is because, to arrive at problem (\ref{eq:optprblm4}), we removed the summand $-1$ from the objective function of (\ref{eq:SDPopennotation}) (and used the property $\max_x a(x) = -\min_x (-a(x))$ to replace the maximization with the minimization). As a result, if $\pi^\ast$ and $p^\ast$ denote the optimal value of problems~(\ref{eq:SDPopennotation}) and (\ref{eq:optprblm4}), respectively, then $\pi^\ast=-1-p^\ast$.\label{foot:objfxn} }
\begin{equation}\label{eq:optprblm4}
\fl \mbox{minimize}\quad \bm{c}^{\sf T} \bm{\lambda}\qquad \mbox{subject to}\quad \bm{F}_0 + \sum_{i=1}^n \bm{F}_i\lambda_i \geq 0
\end{equation}
with
\begin{equation}\label{eq:SDPgivens}
\fl\bm{c}\coloneq-\bm{b}\,,\quad \bm{F}_0\coloneq\left[\begin{array}{c|c}
\begin{array}{c|c}
\bm{Q}_n&\\\hline
&P-1
\end{array}
&\\\hline
&1
\end{array}\right]\quad\mbox{and}\quad 
 \bm{F}_i\coloneq
\left[\begin{array}{c|c}
 \begin{array}{c|c}
 \bm{0}_{n}&\bm{e}_{n,i}\\\hline
\bm{e}_{n,i}^{\sf T}&2
 \end{array}
 &\\\hline
 &-1
 \end{array}\right]\,,
\end{equation}
 where $\bm{\lambda}$, $\bm{b}$, $\bm{Q}_n$ were defined in the~\ref{app:quadtoLMI} and $\bm{e}_{n,i}\in\mathbb{R}^{n\times 1}$ denotes the $n$-dimensional vector whose only non-zero entry is $1$ and occupies the $i$th row. 

The fact that the design of the maximally GM-entangled X-states of $N$-qubits can be cast as a SDP is interesting on its own. First, SDPs are convex optimization problems~\cite{04Boyd} and, as such, have the desirable property that any local optimum is necessarilly a global optimum. Second, many efficient numerical methods have been devised to solve SDPs (for a review, see, e.g., \cite{96Vandenberghe49} and references therein). These methods have excellent convergence properties and output a ``certificate of convergence'', i.e., an interval within which the optimal value of the objective function must lie. Typically, with no more than $30$ iterations, this interval can be made arbitrarily small. Third (and most importantly for our purposes), a powerful duality theory exists for SDPs and can be employed to rigorously prove the optimality of an \emph{ansatz} solution. 

\begin{figure}[h]
\centering
\includegraphics[width=\textwidth]{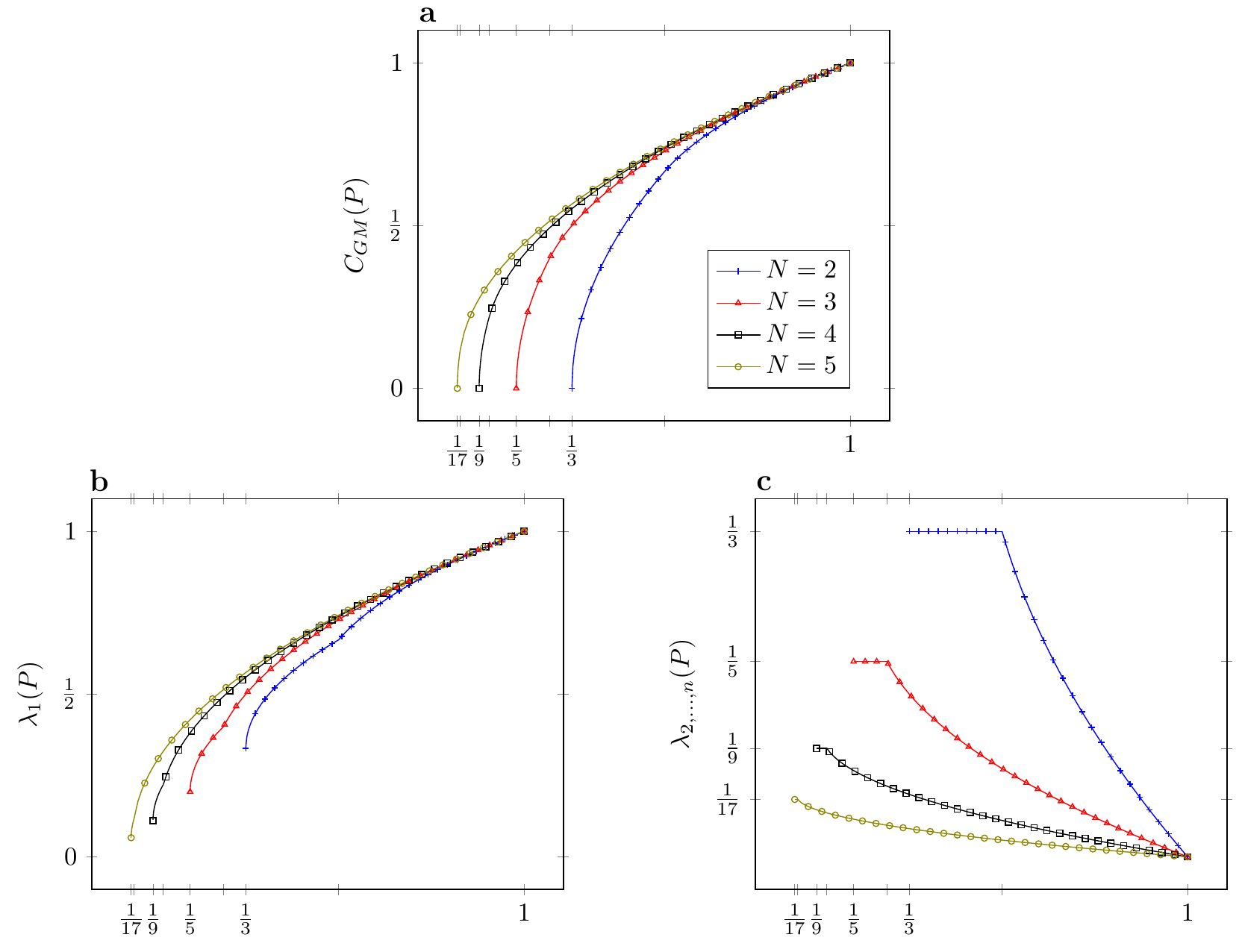}
\caption{Matching between the numerical solution (symbols) of the SDP~(\ref{eq:optprblm4}) and the corresponding analytical formulas (lines) for the optimal GM-concurrence and spectrum  (cf. theorem~\ref{thm:XMEMSwrtPurity}). Each plot considers the parameter values $N=2,3,4,5$ and a uniform sampling of $P\in\,]1/(n+1),1]$ . The legend in plot (a) also applies to plots (b) and (c). The observed agreement between the lines and symbols strongly suggests the optimality of the analytical formulas in the statement of theorem~\ref{thm:XMEMSwrtPurity}. From left to right, the unlabelled tick marks correspond to the purity value $p_n\coloneq(n+3)/(n+1)^2$ for $n=16,8,4,2$ (or, equivalently, $N=5,4,3,2$). Remarkably, although all the plotted functions are continuous, they fail to be smooth: the second derivative of $C_{\rm GM}(P)$ and the first derivatives of $\{\lambda_j(P)\}_{j=1}^n$ are discontinuous at $P=p_n$.}\label{fig:evidopt}
\end{figure}

Before proceeding with our proof, let us exploit the aforementioned numerical virtues of SDPs to provide a first evidence of the optimality of the GM-concurrence $2\gamma$ (cf.~(\ref{eq:gammafxnP})) and of the spectrum~(\ref{eq:propprimalsol}). In figure~\ref{fig:evidopt}, we plot these analytical expressions (lines) along with the numerical solutions (symbols) of problem (\ref{eq:optprblm4}), obtained by running the MATLAB-based solver SeDuMi~\cite{99Sturm625} for several combinations of $P$ and $N$. In our numerical computations, we set SeDuMi's precision to $10^{-15}$, which establishes the largest acceptable length of the aforementioned ``error interval''.

Plot~\ref{fig:evidopt}a shows the converged numerical values of the objective function along with an analytic plot of $2\gamma(P)$. The difference between the numerical and analytical values is found to be of the order of $10^{-14}$, which strongly suggests that $2\gamma(P)$ is the maximal $N$-qubit X-state GM-concurrence wrt purity. Plots~\ref{fig:evidopt}b,c, in turn, indicate the optimality of~(\ref{eq:propprimalsol}). In particular, plot~\ref{fig:evidopt}b reveals an excellent agreement between the converged numerical values of $\lambda_1$ and the function $f(\gamma(P))+\gamma(P)$. Similarly, plot~\ref{fig:evidopt}c illustrates the coincidence between the converged numerical values of $\lambda_{2,\ldots,n}$ (which resulted all the same up to numerical precision) and the function $g(\gamma(P))$, cf. (\ref{eq:propprimalsol}) and (\ref{eq:defsfandg}).

We now briefly review an important result from the SDP duality theory that will be subsequently used to establish the optimality of (\ref{eq:propprimalsol}). For a thorough account on this theory, we refer the reader to~\cite[Chapter 5]{04Boyd}. The \emph{dual problem} of the SDP (\ref{eq:optprblm4}) (henceforth called \emph{primal problem}), is another SDP given by
\begin{equation}\label{eq:dualprblm}
\fl\mbox{maximize}\quad -\tr[{\bm{F}_0 \bm{Z}}]\quad\mbox{subject to}\quad \bm{Z}\geq 0\quad\mbox{and}\quad \tr[{\bm{F}_i\bm{Z}}]=c_i\quad \forall i=1,\ldots,n\,.
\end{equation}
Here, the variable to be optimized is the  matrix $\bm{Z}$, whereas the vector $\bm{c}$ and the matrices $\bm{F}_0$ and $\bm{F}_i$ are the same as the ones appearing in the primal problem (in our particular case, defined in~(\ref{eq:SDPgivens})). Let $p$ denote any feasible value of the primal problem~(\ref{eq:optprblm4}), and denote by $p^\ast$ its optimal value. Similarly, let $d$ and $d^\ast$ denote, respectively, any feasible value and the optimal value of the dual problem~(\ref{eq:dualprblm}). It is obvious that $p\geq p^\ast$ and that $d^\ast\geq d$. Less obvious --- but also true\footnote{To see that, note that the constraints of the primal and dual problems allow us to write:
$$p-d=\bm{c}^{\sf T}\bm{\lambda} +\tr\left[\bm{F}_0\bm{Z}\right]=\sum_{i=1}^n\tr\left[\bm{F}_i \bm{Z}\right]\lambda_i+\tr\left[\bm{F}_0\bm{Z}\right]=\tr\left[\left(\bm{F}_0+\sum_{i=1}^n\bm{F}_i\lambda_i\right)Z\right]\geq 0\,.$$} --- is that $p\geq d$ for \emph{every} primal and dual feasible values (in particular, $p^\ast\geq d^\ast$), in such a way that
\begin{equation}\label{eq:weakduality}
p\geq p^\ast \geq d^\ast\geq d\,,
\end{equation}
a general result known as \emph{weak duality}. Next, we show how these inequalities can be used to prove the optimality of our \emph{ansatz} solution.

As some straightforward computation shows, the spectrum~(\ref{eq:propprimalsol}) can be initially regarded as a \emph{primal feasible point} that yields a \emph{primal feasible value} $p=-1-2\gamma$. However, if we can find a particular \emph{dual feasible point} that yields a \emph{dual feasible value} $d=p$, then weak duality implies that $p=p^\ast$, meaning that~(\ref{eq:propprimalsol}) is, indeed,  a \emph{primal optimal point}. We claim that the block matrix $\bm{\mathcal{Z}}$, described below, provides such dual feasible point:
\begin{equation}\label{eq:matrixZ}
\fl\bm{\mathcal{Z}}=\left[\begin{array}{c|c}
\begin{array}{c|c|c}
\mathfrak{z}_{1}&\sqrt{\mathfrak{z}_1\mathfrak{z}_2}\bm{j}_{n-1}^{\sf T}&-1-\mathfrak{z}_{3}+\frac{\mathfrak{z}_{4}}{2}\\\hline
\sqrt{\mathfrak{z}_1\mathfrak{z}_2}\bm{j}_{n-1}&\mathfrak{z}_{2}\bm{J}_{n-1}&\left(-\frac{1}{2}-\mathfrak{z}_{3}+\frac{\mathfrak{z}_{4}}{2}\right)\bm{j}_{n-1}\\\hline
-1-\mathfrak{z}_{3}+\frac{\mathfrak{z}_{4}}{2}&\left(-\frac{1}{2}-\mathfrak{z}_{3}+\frac{\mathfrak{z}_{4}}{2}\right)\bm{j}_{n-1}^{\sf T}&\mathfrak{z}_{3}
\end{array}
\\\hline
&\mathfrak{z}_{4}
\end{array}\right]\,,
\end{equation}
where $\bm{j}_n\in\mathbb{R}^{n\times 1}$ and $\bm{J}_n\in\mathbb{R}^{n\times n}$ denote the ``all-one'' $n$-dimensional vector and matrix, respectively. In addition,
\begin{equation}\label{eq:z}
\begin{array}{ll}
\fl \mathfrak{z}_{1} \coloneq\left\{\begin{array}{ll}
2+2\gamma+\frac{1}{2\gamma}&\mbox{if}\quad 0 < \gamma\leq \frac{1}{n+1}\\[2mm]
\frac{(1-n)(1+2\gamma)^2}{2(1-2n\gamma)}&\mbox{if}\quad \frac{1}{n+1}\leq \gamma\leq \frac{1}{2}
\end{array}\right.\,,
&\quad \mathfrak{z}_{3}\coloneq\left\{\begin{array}{ll}
\frac{1}{2\gamma}&\mbox{if}\quad 0< \gamma\leq \frac{1}{n+1}\\[2mm]
\frac{1-n}{2(1-2n\gamma)}&\mbox{if}\quad \frac{1}{n+1}\leq \gamma\leq \frac{1}{2}
\end{array}\right.,
\\[10mm]
\fl \mathfrak{z}_{2}\coloneq\left\{\begin{array}{ll}
\frac{(1+\gamma)^2}{2\gamma}&\mbox{if}\quad 0< \gamma\leq \frac{1}{n+1}\\[2mm]
\frac{(n-2\gamma)^2}{2(1-n)(1-2n\gamma)}&\mbox{if}\quad \frac{1}{n+1}\leq \gamma\leq \frac{1}{2}
\end{array}\right.\,,
&\quad \mathfrak{z}_{4}\coloneq \left\{\begin{array}{ll}
0&\mbox{if}\quad 0< \gamma\leq \frac{1}{n+1}\\[2mm]
1+\frac{1-2\gamma}{1-2n\gamma}&\mbox{if}\quad \frac{1}{n+1}\leq \gamma\leq \frac{1}{2}
\end{array}\right.\,. 
\end{array}
\end{equation}

To verify our claim, we first show that $\bm{\mathcal{Z}}$ establishes $d=p$. Indeed, $d=-\tr\left[\bm{F}_0\bm{\mathcal{Z}}\right]$, which can be evaluated with the aid of (\ref{eq:SDPgivens}) and (\ref{eq:matrixZ}) to give
\begin{equation}
d=-\frac{1}{n+1}\left[\mathfrak{z}_1 n + 2(n-1)(\mathfrak{z}_2-\sqrt{\mathfrak{z}_1 \mathfrak{z}_2})\right]-\mathfrak{z}_3(P-1)-\mathfrak{z}_4\,.\label{eq:dualvalueZ}\end{equation}
Then, plugging (\ref{eq:z}) into (\ref{eq:dualvalueZ}) and expressing $P$ as the sum of the squares of the eigenvalues in~(\ref{eq:propprimalsol}), we obtain $d=-1-2\gamma = p$, as claimed.

Now, we show that $\bm{\mathcal{Z}}$ satisfies the constraints of problem (\ref{eq:dualprblm}). Regardless of the values $\mathfrak{z}_{1,2,3,4}$, it is easy to check that $\tr\left[\bm{F}_1 \bm{\mathcal{Z}}\right]=-2$ and $\tr\left[\bm{F}_{2,\ldots,n} \bm{\mathcal{Z}}\right]=-1$, as required. Finally, the condition $\bm{\mathcal{Z}}\geq 0$ can be checked by an explicit study of the eigenvalues of $\bm{\mathcal{Z}}$. Simple inspection of (\ref{eq:z}) shows that $\mathfrak{z}_4\geq 0$. Furthermore, with some cumbersome simplification procedure, the characteristic polynomial of the first $(n+1)$-dimensional block of $\bm{\mathcal{Z}}$ takes the form
\begin{equation}
x^{n+1}-\Lambda_n(\gamma)x^n=0
\end{equation}
with
\begin{equation}\label{eq:Lambdan}
\Lambda_n(\gamma)\coloneq\left\{\begin{array}{cl}
\frac{(n+1)(1+2\gamma)+(n+3)\gamma^2}{2\gamma}&\mbox{if}\quad 0< \gamma\leq \frac{1}{n+1}\\[2mm]
\frac{n^2+2(n-1)-4\gamma+4n\gamma^2}{2(-1+2n\gamma)}&\mbox{if}\quad \frac{1}{n+1}\leq \gamma\leq \frac{1}{2}
\end{array}\right.\,,
\end{equation}
thus establishing $x=\Lambda_n(\gamma)$ as the only non-zero eigenvalue of this block. Clearly, the first branch of (\ref{eq:Lambdan}) is strictly positive. To see that this is also true for the second branch, define
\begin{equation}
\fl F_n(\gamma)\coloneq n^2+2(n-1)-4\gamma+4n\gamma^2\quad\mbox{and}\quad G_n(\gamma)\coloneq {2(-1+2n\gamma)}
\end{equation}
in such a way that, for $\gamma\in[1/(n+1),1/2]$,  $\Lambda_n(\gamma)=F_n(\gamma)/G_n(\gamma)$. We conclude our proof by showing that, for $\gamma\geq 1/(n+1)$, both $F_n(\gamma)$ and $G_n(\gamma)$ are strictly positive, hence so is $\Lambda_n(\gamma)$. To establish the positivity of $G_n(\gamma)$, note the following implications:
\begin{equation}
\gamma\geq \frac{1}{n+1} \Rightarrow \gamma > \frac{1}{2n} \Rightarrow G_n(\gamma) >0\,.
\end{equation}
The first implication follows from the fact that, for the relevant values of $n$ (recall that $n\geq 2$), the inequality $n+1<2n$ holds trivially. The positivity of $F_n(\gamma)$, in turn, can be established by noting that
\begin{equation}
\fl F_n\left(\frac{1}{n+1}\right)=\frac{(n-1)(n+3)[2+n(n+2)]}{(n+1)^2}>0\quad\mbox{and}\quad \frac{\partial F_n(\gamma)}{\partial \gamma} = 2 G_n(\gamma)>0\,.
\end{equation}
So, we see that $F_n(\gamma)$ is already positive at $\gamma=1/(n+1)$ and monotonically increasing for $\gamma>1/(n+1)$.

\section{Converting between density matrices with few Kraus operators}\label{sec:convert}

For any given pair of isospectral density matrices, it is possible to find a unitary transformation that maps one density matrix into the other. In Theorem~\ref{thm:XMEMSwrtSpectrum}, for example, all unitary transformations that map an arbitrary $N$-qubit density matrix into a corresponding X-MEMS wrt spectrum were explicitly constructed. However, if two density matrices have different spectra (e.g., an arbitrary density matrix and a corresponding X-MEMS wrt purity, cf. Sec.~\ref{sec:XMEMSwrtPurity}), then there is no unitary map capable of converting between the two, in which case one must resort to more general quantum operations to implement the desired state transformation. In this section, we introduce a numerical scheme to design a quantum channel $\bm{\mathcal{C}}$, as modelled by a completely positive and trace preserving (CPTP) map, that promotes the conversion between any two given density matrices. Furthermore, our scheme constrains the resulting map to be ``economical'' wrt certain resources utilized in its implementation.

To understand what \emph{economical} means in this context, consider the general representation of $\bm{\mathcal{C}}$ in terms of Kraus operators $\bm{M}_m\in\mathcal{B}({\sf H}_d)$:
\begin{equation}
\bm{\mathcal{C}}(\bm{\rho})=\sum_m \bm{M}_m \bm{\rho} \bm{M}_m^\dagger\quad\mbox{with}\quad \sum_m\bm{M}_m^\dagger\bm{M}_m=\Id\,.
\end{equation}
A trivial choice of $\bm{\mathcal{C}}$, such that $\overline{\bm{\rho}}=\bm{\mathcal{C}}(\bm{\rho})$, is the channel that maps \emph{every} density matrix to $\overline{\bm{\rho}}$, which admits the following \emph{minimal} set of Kraus operators:
\begin{equation}\label{eq:sillyMm}
\{\bm{M}_m\}_{m=1}^{d\mathfrak{r}}=\{\sqrt{a_\mu}\ket{\mu}\!\bra{\nu}\}_
{\mu=1,\ldots,\mathfrak{r}\atop\nu=1,\ldots,d}\,,
\end{equation}
where $d\coloneq \dim\overline{\bm{\rho}}$, $\mathfrak{r}\coloneq\rnk\overline{\bm{\rho}}$, $\{\ket{\nu}\}_{\nu=1}^d$ is an orthonormal basis formed from the eigenvectors of $\overline{\bm{\rho}}$  and $\{a_\mu\}_{\mu=1}^{\mathfrak{r}}$ are the non-zero eigenvalues of $\overline{\bm{\rho}}$ (corresponding to the eigenvectors $\ket{\mu}$), i.e.,
\begin{equation}
\overline{\bm{\rho}}=\sum_{\mu=1}^{\mathfrak{r}}a_\mu \ket{\mu}\!\bra{\mu}\,.
\end{equation}
Practically, though, implementing~(\ref{eq:sillyMm}) can be considered an overkill. In fact, since we \emph{only} require $\bm{\rho}\mapsto\overline{\bm{\rho}}$, it might be possible to find a quantum channel that implements the desired state transformation with a number of Kraus operators much smaller than $d\mathfrak{r}$. In other words, there might be a more \emph{economical CPTP map} for the state transformation $\bm{\rho}\mapsto\overline{\bm{\rho}}$.

With that mind-set, consider the task of determining, amongst every CPTP map $\bm{\mathcal{C}}$ that satisfies $\bm{\mathcal{C}}(\bm{\rho})=\overline{\bm{\rho}}$, those that can be decomposed with the smallest possible number of Kraus operators. Mathematically, this leads to an optimization problem that can be nicely expressed with the aid of the Choi-Jamio{\l}koswki isomorphism~\cite{72Jamiolkowski275,75Choi285,99Fujiwara3290,99Horodecki1888}, which brings CPTP maps $\bm{\mathcal{C}}: \mathcal{B}({\sf H}_d)\to \mathcal{B}({\sf H}_d)$  into a one-to-one correspondence with positive semidefinite matrices $\bm{\mathfrak{C}}\in\mathcal{B}({\sf H}_d\otimes {\sf H}_d)$ such that $\tr_2[\bm{\mathfrak{C}}]=\Id_d$ (here and throughout, $\tr_x$ denotes the partial trace over the $x$th $d$-dimensional subsystem). In particular, given a CPTP map $\bm{\mathcal{C}}: \mathcal{B}({\sf H}_d)\to \mathcal{B}({\sf H}_d)$, its corresponding Choi-Jamio{\l}koswki matrix is
\begin{equation}
\bm{\mathfrak{C}}=(\bm{\mathcal{I}}\otimes\bm{\mathcal{C}})\ket{\Psi}\!\bra{\Psi}\,,
\end{equation}
where $\bm{\mathcal{I}}$ is the identity map on $\mathcal{B}({\sf H}_d)$ and $\ket{\Psi}$ is the (unnormalized) maximally entangled state 
$\ket{\Psi}=\sum_{\alpha=1}^d\ket{h_d^\alpha}\otimes\ket{h_d^\alpha}$, with $\{\ket{h_d^\alpha}\}_{\alpha=1}^d$  a fixed orthonormal basis for ${\sf H}_d$. In this framework, the minimal Kraus decompositions of $\bm{\mathcal{C}}$ can be shown to have $\rnk\bm{\mathfrak{C}}$ Kraus operators~\cite{02Verstraete}, and the equation $\bm{\mathcal{C}}(\bm{\rho})=\overline{\bm{\rho}}$ is equivalent to $\tr_1\left[(\bm{\rho}^{\sf T}\otimes \Id_d)\bm{\mathfrak{C}}\right]=\overline{\bm{\rho}}$ (the transposition being taken wrt $\bm{\rho}$ written in the basis $\{\ket{h_d^\alpha}\}_{\alpha=1}^d$)~\cite{01DAriano042308}. Accordingly, the optimization problem takes the form
\begin{eqnarray}
\mbox{minimize}&\quad\rnk{\bm{\mathfrak{C}}}\nonumber\\
\mbox{subject to}&\quad \bm{\mathfrak{C}}\geq 0\,,\quad \tr_2[\bm{\mathfrak{C}}]=\Id_d\quad\mbox{and}\quad \tr_1\left[(\bm{\rho}^{\sf T}\otimes \Id_d)\bm{\mathfrak{C}}\right]=\overline{\bm{\rho}}\,.\label{eq:rnkproblem}
\end{eqnarray}

Problem~(\ref{eq:rnkproblem}) is an example of rank minimization problem (RMP) with SDP constraints. Although special cases of this problem have been solved (see, e.g.,~\cite{10Recht471} and references therein), RMP are, in general, computationally intractable (NP-hard) due to the non-smoothness and non-convexity of the rank function. Fortunately, though, heuristic methods exist to efficiently approximate their solutions~\cite{02Fazel,01Fazel4734,03Fazel2156}. Essentially, these heuristics rely on replacing the rank with a surrogate function, in such a way that the resulting problem can be handled with standard SDP solvers. Next, we consider the application of two such methods to problem~(\ref{eq:rnkproblem}).

\subsection{Trace heuristic}
Basically, it consists of replacing $\rnk\bm{\mathfrak{C}}$ with $\tr[\bm{\mathfrak{C}}]$. Intuitively, the replacement makes sense because sparse vectors tend to have a $\ell_1$-norm smaller than dense vectors. So, by forming a vector from the eigenvalues of $\bm{\mathfrak{C}}$ and minimizing its $\ell_1$-norm (which corresponds to minimize $\tr[\bm{\mathfrak{C}}]$ when $\bm{\mathfrak{C}}\geq 0$), we may be effectively vanishing some eigenvalues of $\bm{\mathfrak{C}}$ and, hence, reducing its rank (see~\cite{01Fazel4734,02Fazel} for a more technical justification of the trace heuristic in terms of the convex envelope of the rank). To a large extent, the appeal of the trace heuristic comes from the fact that it  provides a \emph{linear} objective function for the optimization problem, which usually means that the solution can be efficiently obtained (at least numerically). Unfortunately, in the case of problem~(\ref{eq:rnkproblem}), the constraint $\tr_2[\bm{\mathfrak{C}}]=\Id_d$ implies that $\tr[\bm{\mathfrak{C}}]=d$, so no minimization can actually occur. As a result, the trace heuristic is useless for our purposes.

\subsection{Log-det heuristic}
Introduced in~\cite{02Fazel,03Fazel2156}, the log-det heuristic can be considered a refinement of the trace heuristic. Consists of replacing $\rnk\bm{\mathfrak{C}}$ with $\log\det(\bm{\mathfrak{C}}+\delta\Id_{d^2})$, where $\delta>0$ is a regularization constant. The value of $\delta$ can be made arbitrarily small, and it is used to avoid an ill-defined objective function as $\bm{\mathfrak{C}}$ gets singular ($\det{\bm{\mathfrak{C}}}=0$). Intuitively, it is expected that, for $\bm{\mathfrak{C}}\geq 0$, the determinant $\det(\bm{\mathfrak{C}}+\delta \Id_{d^2})$ will decrease as the eigenvalues of $\bm{\mathfrak{C}}$ vanish. So, we attempt to minimize the function $\log\det(\bm{\mathfrak{C}}+\delta \Id_{d^2})$, which plays the role of a smooth and concave~\cite{04Boyd} surrogate of the rank.

The optimization problem resulting from the application of the log-det heuristic to~(\ref{eq:rnkproblem}) is a minimization of a concave function over a convex set, and it is thus non-convex.  In order to obtain a related convex optimization problem, the log-det objective function is expanded (to the first order) in a Taylor series about a fixed $d^2$-dimensional matrix $\bm{\mathfrak{C}}_i$, in such a way as to obtain the linear approximation
\begin{equation}
\log\det(\bm{\mathfrak{C}}+\delta \Id_{d^2})\approx\log\det(\bm{\mathfrak{C}}_i+\delta \Id_{d^2})+\tr\left[(\bm{\mathfrak{C}}_i+\delta\Id_{d^2})^{-1}(\bm{\mathfrak{C}}-\bm{\mathfrak{C}}_i)\right]
\end{equation}
where we have used that, for $\bm{X}>0$, $\nabla_{\bm{X}}\log\det \bm{X} = \bm{X}^{-1}$ \cite[pp. 641,642]{04Boyd}. Then, dropping the (irrelevant) constant terms, we end up with the SDP
\begin{eqnarray}
\mbox{minimize}&\quad \tr\left[(\bm{\mathfrak{C}}_i+\delta\Id_{d^2})^{-1}\bm{\mathfrak{C}}\right]\nonumber\\
\mbox{subject to}&\quad \bm{\mathfrak{C}}\geq 0\,,\quad \tr_2[\bm{\mathfrak{C}}]=\Id_d\quad\mbox{and}\quad \tr_1\left[(\bm{\rho}^{\sf T}\otimes \Id_d)\bm{\mathfrak{C}}\right]=\overline{\bm{\rho}}\,.\label{eq:logdetheuristicproblemlinear}
\end{eqnarray}

The minimum of $\log\det(\bm{\mathfrak{C}}+\delta \Id_{d^2})$ can be approximated by iteratively solving the SDP (\ref{eq:logdetheuristicproblemlinear}), taking for $\bm{\mathfrak{C}}_{i+1}$ the resulting $\bm{\mathfrak{C}}$ of the problem solved with input matrix $\bm{\mathfrak{C}}_i$. If we set $\bm{\mathfrak{C}}_0=(1-\delta)\Id_{d^2}$ as the input for the first iteration, the resulting optimization problem coincides with that obtained with the trace heuristic, in which case a rank reduction may not occur in the first iteration. For successive steps, though, rank reduction is generally observed, justifying the initial claim that the log-det heuristic is a refinement of the trace heuristic.

\subsection{Application: Producing X-MEMS with respect to purity}

As an illustration of the effectiveness of the log-det heuristic, let us design economical CPTP maps to transform a given input state of purity $P$ into an X-MEMS wrt $P$, denoted by $\bm{\rho}_P^\prime$ (density matrix obtained by plugging (\ref{eq:gammafxnP}), (\ref{eq:propprimalsol}) and (\ref{eq:defsfandg}) into (\ref{eq:rhoxprime})). For the input state, we take the following $N$-qubit density matrix of purity $P=1/(N+1)$:
\begin{equation}
\bm{\rho}_N=\frac{1}{N+1}\sum_{k=0}^N \ket{D_k^N}\!\bra{D_k^N}\,,
\end{equation}
where $\ket{D_k^N}$ are the (totally symmetric) $N$-qubit Dicke states of $k$ excitations~\cite{54Dicke99,03Stockton022112}, defined in the computational basis as
\begin{equation}
\ket{D_k^N}\coloneq\sqrt{\frac{k! (N-k)!}{N!}}\sum_{\sigma}\ket{\sigma(1,\stackrel{k}{\cdots},1,0,\stackrel{N-k}{\cdots},0)}\,,
\end{equation}
with the summation running over every distinct permutation of the sequence of $k$ ones and $N-k$ zeros. 

\begin{figure}[h]
\centering
\includegraphics[width=\textwidth]{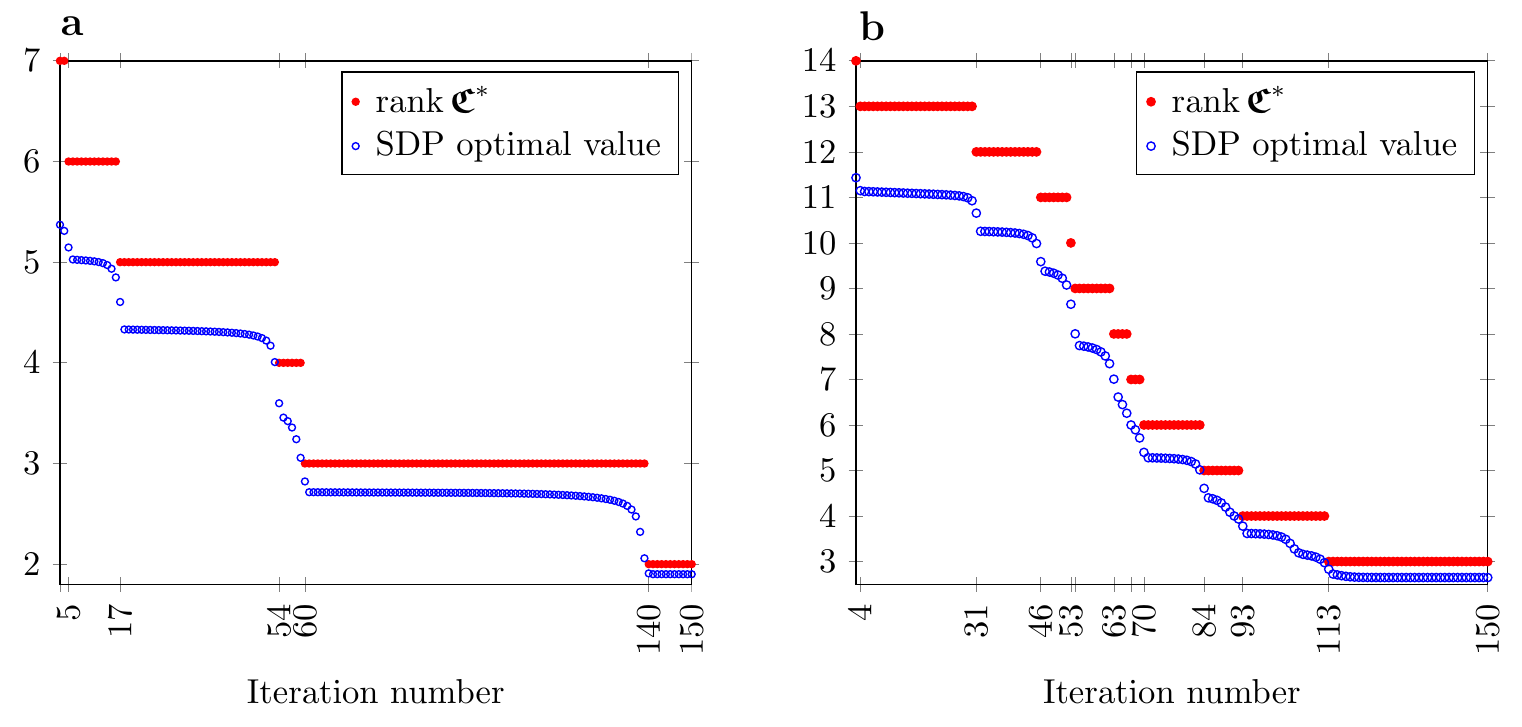}
 \caption{Rank decay of the Choi-Jamio{\l}kowski matrix as the iterations of the log-det heuristic progress with parameters $\delta=0.2$ and $\bm{\mathfrak{C}}_0=\Id_d\otimes \bm{\rho}_P^\prime$. Plots (a) and (b) correspond to the cases $N=3$ and $N=4$, respectively. Tick marks are shown on the iterations where a rank drop takes place. For ease of visualization, the optimal values of the rank of $\bm{\mathfrak{C}}$ and of the objective function of~(\ref{eq:logdetheuristicproblemlinear}) are presented from the third iteration onwards. For plot (a), the rank of the initial Choi-Jamio{\l}kowski matrix is $d\mathfrak{r}=8\times 5=40$, which decays to $24$ during the first and second iterations and reaches the minimum $2$ after $140$ iterations. For plot (b), the initial rank is $d\mathfrak{r}=16\times 8=128$, decaying to $51$ during the first and second iterations. After $113$ iterations the rank reaches $3$ and no further decay is observed.}\label{fig:rankdrop}
\end{figure}

Using SeDuMi, many iterations of the SDP (\ref{eq:logdetheuristicproblemlinear}) are solved to produce a low rank Choi-Jamio{\l}kowski matrix that maps $\bm{\rho}=\bm{\rho}_N$ to $\overline{\bm{\rho}}=\bm{\rho}^\prime_P$. Figure~\ref{fig:rankdrop} shows the evolution of $\rnk\bm{\mathfrak{C}}^\ast$ as the iterations progress, where $\bm{\mathfrak{C}}^\ast$ denotes the optimal Choi-Jamio{\l}kowski matrix found at each step. For completeness and comparison, we also plot the obtained optimal values of the SDP~(\ref{eq:logdetheuristicproblemlinear}) at each iteration. Our numerical simulations were run with $\delta=0.2$ and $\bm{\mathfrak{C}}_0=\Id_d\otimes \bm{\rho}_P^\prime$, which corresponds to the CPTP map that collapses the entire Bloch ball into the point corresponding to $\bm{\rho}_P^\prime$, cf. (\ref{eq:sillyMm}). Plots~\ref{fig:rankdrop}a and \ref{fig:rankdrop}b correspond to the cases $N=3$ and $N=4$, respectively.

In the case $N=3$, plot~\ref{fig:rankdrop}a shows that $\rnk{\bm{\mathfrak{C}}^\ast}=2$ is reached  on the $140$th iteration. From there onwards, every produced $\bm{\mathfrak{C}}^*$ determines a CPTP map that can be written with only two Kraus operators\footnote{See, e.g.,~\cite[Appendix B]{11Fanchini677} for a summary on how to construct sets of Kraus operators for a CPTP map from its Choi-Jamio{\l}kowski matrix.}. Remarkably, in this case, the log-det heuristic provides \emph{minimal} Kraus decompositions, since no further decrease can occur (as $\rnk{\bm{\mathfrak{C}}^\ast}=1$ would correspond to a unitary map). In the case $N=4$, plot~\ref{fig:rankdrop}b shows that $\rnk{\bm{\mathfrak{C}}^\ast}=3$ is reached after $113$ iterations. Although not shown in the plot, we ran $150$ iterations more and, even so, no further decrease of $\rnk{\bm{\mathfrak{C}}^\ast}$ was observed. Of course, that does not mean that a CPTP map that implement $\bm{\rho}_4\mapsto\bm{\rho}_{1/5}^\prime$ with two Kraus operators does not exist, but merely indicates that if it does exist, then it cannot be reached with the log-det heuristic. 

We conclude this section by noting that, although the log-det heuristic may not always lead to minimal Kraus decompositions, it is at least very effective in producing CPTP maps that implement a desired state transformation with a relatively small number of Kraus operators. It should be noted, though, that the efficiency of the rank decay scheme is significantly limited by the number of qubits involved, as the size of problem~(\ref{eq:logdetheuristicproblemlinear}) scales exponentially with $N$.

\section{Concluding Remarks}\label{sec:conclusion}
The characterization of multiqubit MEMS is a long-standing open problem in quantum information science. In its core, lies the inherent difficulty in quantifying the genuine multipartite entanglement of multiqubit systems~\cite{09Guhne1,09Horodecki865}. Notwithstanding, as progresses start to be made in this field~\cite{10Huber210501,11Ma062325,12Rafsanjani062303}, some preliminar sketches of how $N$-qubit MEMS look like can be drawn~\cite{13Agarwal1350043}. In this paper, we rely on a recently obtained closed formula for the GM-concurrence of $N$-qubit $X$-states~\cite{12Rafsanjani062303} to determine --- amongst the set of $N$-qubit X-states --- those with maximal GM-concurrence for (i) a fixed set of eigenvalues and (ii) for a fixed mixedness (as measured by purity), which we refer to as ``X-MEMS wrt spectrum'' and ``X-MEMS wrt purity'', respectively.

Using only elementary algebra, explicit forms of density matrices and maximal GM-concurrence were obtained for X-MEMS wrt every possible spectrum. Besides, the unitary transformation that takes an arbitrary $N$-qubit state into the corresponding X-MEMS wrt spectrum was characterized, generalizing a previous result of Verstraete \emph{et al.}~\cite{01Verstraete12316} for two-qubits. Although X-MEMS wrt purity had already been identified in~\cite{13Agarwal1350043}, we relied on the fact that they form a subset of X-MEMS wrt to spectrum to numerically reconstruct them and rigorously prove their optimality via SDP methods. Additionally, we formulated as a rank minimization problem the design of quantum operations that implements any desired quantum state transformation with the minimal number of Kraus operators. Then, applying a heuristic method to specific examples of this optimization problem with $N=3$ and $4$ qubits, we efficiently characterized low rank quantum operations that transform three- and four-qubit states into the corresponding X-MEMS wrt purity.

An extension of our SDP approach to characterize extreme X-states wrt other measures of mixedness (e.g. von Neumann entropy) and/or other measures of multipartite quantum correlations/nonclassicality (e.g. GM-negativity~\cite{11Jungnitsch190502}, global quantum discord~\cite{11Rulli042109} or the measure introduced in~\cite{11Giorgi190501}) is an interesting line for future research. For any desired measures of correlation ($\mathfrak{c}$) and mixedness ($\mathfrak{m}$), the corresponding extreme X-states are, formally, the optimal points of the problem
\begin{equation}\label{eq:optprblmgen}
\fl \mbox{maximize}\quad\mathfrak{c}(\bm{\rho}_X)\quad\mbox{subject to}\quad \bm{\rho}_X\geq 0\,,\quad \tr\bm{\rho}_X=1\,,\quad \mathfrak{m}(\bm{\rho}_X)=\mathfrak{m}_0\,,
\end{equation}
where $\mathfrak{m}_0$ specifies a desired value of mixedness and the optimization runs over all X-density matrices $\bm{\rho}_X$. Although linearity of $\mathfrak{c}$ and $\mathfrak{m}$ would promptly guarantee that problem (\ref{eq:optprblmgen}) is a SDP, such a form can also be established in certain non-linear cases. Indeed, in this paper, we have seen that despite the specific non-linearities of $\mathfrak{c}$ (taken as the GM-concurrence) and $\mathfrak{m}$ (taken as the purity), an equivalent SDP was built by suitably parametrizing $\mathfrak{c}(\bm{\rho}_X)$ (cf. (\ref{eq:eigsCgeNqubitXstate})) and applying a standard trick to turn  $\mathfrak{m}(\bm{\rho}_X)=\mathfrak{m_0}$ into a LMI (cf.~\ref{app:quadtoLMI}). It is thus conceivable that a similar approach can handle other choices of non-linear measures $\mathfrak{c}$ and $\mathfrak{m}$. However, determining whether (and how) problem (\ref{eq:optprblmgen}) can be cast as a SDP is expected to strongly depend on the particular choice of measures and, as such, should be considered case by case. 

Along these lines, a particularly interesting problem is to consider a pair of optimization problems formed by fixing $\mathfrak{c}$ as some correlation measure and setting $\mathfrak{m}$ as (i) purity and (ii) von Neumman entropy. The question to be answered here is whether the two problems yield the same set of extreme X-states. Such an analysis has been conducted by Wei \emph{et al.} in the bipartite case with $\mathfrak{c}$ set as concurrence, negativity and relative entropy of entanglement~\cite{03Wei22110}. Remarkably, different extreme states were obtained by changing the mixedness measure, which suggests that the same would occur in the multipartite setting. However, an explicit verification of this conjecture remains an open problem.

\ack
The authors are indebted to Marcus Huber and Mart\'i Perarnau-Llobet for bringing reference~\cite{14Huber} to our attention, and to an anonymous referee for valuable comments on a previous version of this manuscript. PEMFM thanks G J Milburn for discussions and the financial support jointly provided by the Brazilian Air Force and by the program ``Ci\^encia sem Fronteiras'', Project No. 200024/2014-0. SMHR acknowledges financial support from National Science Foundation Grant No. PHY-1203931.

\appendix
\section{Converting a quadratic inequality into a linear matrix inequality}\label{app:quadtoLMI}
In this appendix we briefly review a standard trick to convert quadratic inequality constraints into linear matrix inequalities (LMI). In our proof of Theorem~\ref{thm:XMEMSwrtPurity}, this was used to write the (convex) quadratically constrained linear program~(\ref{eq:optprblm3}) in the SDP inequality form~(\ref{eq:optprblm4}). We illustrate the technique in this particular case.

First, we use the well-known formula
\begin{equation}
\left(\sum_{k=1}^n x_k\right)^2=\sum_{k=1}^n x_k^2+ 2\sum_{\ell>k=1}^n x_k x_\ell\,,
\end{equation}
to rewrite the quadratic constraint in~(\ref{eq:optprblm3}) as
\begin{equation}
P-1+2\sum_{k=1}^n\lambda_k-2\sum_{\ell\geq k=1}^n\lambda_k\lambda_\ell\geq 0
\end{equation}
or, equivalently,
\begin{equation}\label{eq:quadform}
P-1+2\bm{j}_n^{\sf T}\bm{\lambda}-\bm{\lambda}^{\sf T}\bm{Q}_n^{-1}\bm{\lambda}\geq 0\,,
\end{equation}
where $\bm{j}_n\in\mathbb{R}^{n\times 1}$ and $\bm{J}_n\in\mathbb{R}^{n\times n}$ denote the ``all-one'' $n$-dimensional vector and matrix, respectively, and
\begin{equation}
\bm{\lambda}\coloneq(\lambda_1\,,\,\lambda_2\,,\ldots\,,\,\lambda_n)^{\sf T}\quad\mbox{and}\quad \bm{Q}_n^{-1}\coloneq \Id_n + \bm{J}_n\,.
\end{equation}
We note that $\bm{Q}_n^{-1}$ is a positive non-singular matrix with eigenvalues $1$ ($(n-1)$-fold degenerate) and $n+1$ (non-degenerate), and its inverse is given by
\begin{equation}
\bm{Q}_n=\Id_n-\frac{1}{n+1}\bm{J}_n\,.
\end{equation}

Now, we recognize the lhs of inequality~(\ref{eq:quadform}) as the Schur complement of the following (block) matrix of dimension $n+1$:
\begin{equation}\label{eq:schurcompl}
\left[\begin{array}{c|c}
\bm{Q}_n&\bm{\lambda}\\\hline
\bm{\lambda}^{\sf T}&P-1+2\bm{j}_n^{\sf T}\bm{\lambda}
\end{array}\right]\,.
\end{equation}
Since $\bm{Q}_n>0$, we conclude that~(\ref{eq:quadform}) is equivalent to the constraint that matrix~(\ref{eq:schurcompl}) is positive semidefinite~\cite[pp. 650-651]{04Boyd}. As a result, problem~(\ref{eq:optprblm3}) takes the form:

\begin{equation}\label{eq:SDPopennotation}
\fl \mbox{maximize}\quad -1+\bm{b}^{\sf T}\bm{\lambda}\qquad
\mbox{subject to}\quad
\left[\begin{array}{c|c}
\begin{array}{c|c}
\bm{Q}_n&\bm{\lambda}\\\hline
\bm{\lambda}^{\sf T}&P-1+2\bm{j}_n^{\sf T}\bm{\lambda}
\end{array}
& 
\\\hline
&1-\bm{j}_n^{\sf T}\bm{\lambda}
\end{array}\right]\geq 0
\end{equation}
where $\bm{b}\coloneq(2\,,\,1\,,\,\ldots\,,\,1)^{\sf T}$. Up to a constant term and an inversion of sign in the objective function (see footnote at page~\pageref{foot:objfxn}), this is equivalent to problem~(\ref{eq:optprblm4}).

\section*{References}

\providecommand{\newblock}{}

\end{document}